\documentclass[dvips]{article}
\usepackage[belowskip=-25pt,aboveskip=0pt]{caption}
\usepackage{paralist}
\usepackage{lineno}
\usepackage[compact]{titlesec}
\usepackage{icrc2011}

\title{LOFAR: Detecting Cosmic Rays with a Radio Telescope}

\newcommand{\etal}{\MakeLowercase{\textit{et al.}}} % "et al."
\shorttitle{A.~Corstanje \etal, LOFAR}

\authors{
  A.~Corstanje$^{1}$, 
  M.~van den Akker$^{1}$,
  L.~B\"ahren$^{1,4}$,
  H.~Falcke$^{1,2}$,
  W.~Frieswijk$^{2}$,
  J.~R.~H\"orandel$^{1}$,
  A.~Horneffer$^{1,5}$,
  C.~W.~James$^{1,6}$,
  J.~L.~Kelley$^{1}$,
  R.~McFadden$^{2}$, 
  M.~Mevius$^{3}$,
  A.~Nelles$^{1}$,
  P.~Schellart$^{1}$,
  O.~Scholten$^{3}$,
  S.~Thoudam$^{1}$,
  S.~ter Veen$^{1}$}

\afiliations{$^1$Department of Astrophysics, IMAPP, Radboud University
  Nijmegen, 6500GL Nijmegen, The Netherlands\\
 $^2$Netherlands Institute for Radio Astronomy (ASTRON), 7990AA Dwingeloo,
  The Netherlands\\
 $^3$Kernfysisch Versneller Instituut, 9747AA Groningen, The Netherlands\\
 $^4$now at: Anton Pannekoek Astronomical Institute, University of Amsterdam,
  1090 GE Amsterdam, The Netherlands\\
 $^5$now at: Max-Planck-Institut f\"{u}r Radioastronomie, 53121 Bonn,
  Germany\\
 $^6$now at: Erlangen Centre for Astroparticle Physics, University of
  Erlangen-Nuremberg, 91058 Erlangen, Germany
}

\email{a.corstanje@astro.ru.nl}

\abstract{LOFAR (the Low Frequency Array), a distributed digital
  radio telescope with stations in the Netherlands, Germany, France, Sweden,
  and the United Kingdom, is designed to enable full-sky monitoring of
  transient radio sources.  These capabilities are ideal for the detection
  of broadband radio pulses generated in cosmic ray air showers. The core
  of LOFAR consists of 24 stations within 4 square 
  kilometers, and each station contains 96 low-band antennas and 48
  high-band antennas.  This dense instrumentation will allow detailed
  studies of the lateral distribution of the radio signal in a frequency
  range of 10-250 MHz.  Such studies are key to understanding the various
  radio emission mechanisms within the air shower, as well as for
  determining the potential of the radio technique for primary particle
  identification. We present the status of the LOFAR cosmic ray program,
  including the station design and hardware, the triggering and filtering
  schemes, and our initial observations of cosmic-ray-induced radio pulses.}


\begin{document}
\maketitle

% LINE NUMBERS
%\modulolinenumbers[5]
%\setlength\linenumbersep{3pt}
%\linenumbers

%------------------------------------------------------------

\section{Introduction}

Radio emission from cosmic ray air showers was discovered in 1965 by Jelley
and collaborators \cite{jelley_radio}.  Advances in high-speed digital
electronics have expanded the options for filtering, triggering, and
recording the radio signal and have led to renewed interest in the technique.
This window on high-energy cosmic rays has a duty cycle of
nearly 100\% and can in principle, like the air fluorescence technique,
provide information on shower development in the atmosphere.

The LOFAR Prototype Station (LOPES) successfully detected the
coherent radio pulses from air showers with energies from $5 \times 10^{16}$
to $5 \times 10^{17}$ eV in a frequency range of 43-73 MHz.  LOPES verified
the primary nature of the emission from the 
air shower: the interaction of shower electrons and positrons with the
Earth's magnetic field \cite{falcke_nature}.  Since then, significant
progress has also been made in the theoretical
understanding of the radio signals \cite{reas3_mgmr}, with current work focused
on sub-dominant emission mechanisms, such as charge excess
and Cherenkov contributions.  

LOFAR, the Low Frequency
Array, is a digital radio telescope with a dense core of antennas capable
of detecting air shower pulses and recording the signal polarization and lateral
distribution with unprecedented accuracy.  Understanding and modeling all
details of the radio emission will allow a full demonstration of the
potential for determining fundamental cosmic ray properties from the radio
signal, such as primary particle type \cite{mgmr_composition, huege_composition}.  

%------------------------------------------------------------

\section{LOFAR}

\begin{figure}
%\centerline{ \includegraphics[width=0.4 \textwidth]{superterp_sml.eps}}
\centering
\includegraphics[width=0.4 \textwidth]{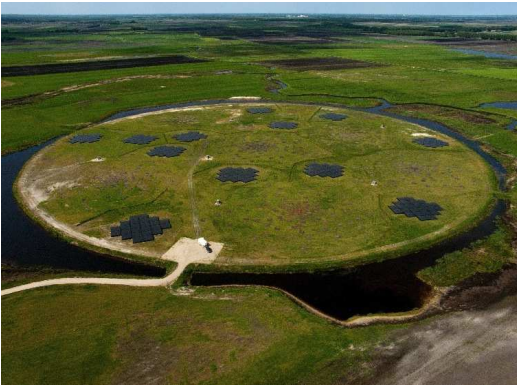}
\caption{Photograph of the LOFAR superterp -- pictured are twelve HBA
core fields (grey squares) and six LBA fields (barely visible).} \label{superterp}
\end{figure}

LOFAR is a digital aperture-synthesis radio telescope
consisting of at least 48 stations. $24$ core stations are concentrated within
a 4-$\mathrm{km}^2$ area near Exloo in the Netherlands, 16 remote stations are distributed
up to $65$~km from the core throughout the northern Netherlands, and 8 international
stations are spread between England, France, Germany, and Sweden.
Unlike typical ground-based cosmic ray detectors, the density of LOFAR stations
from the core falls off smoothly, from six stations in the ``superterp'' (the
$360$~m diameter island at the center of the LOFAR core; see
Fig.~\ref{superterp}), to effectively isolated remote and international stations.

%\begin{figure}
%%\centerline{ \includegraphics[width=0.4 \textwidth]{lba.eps}}
%\centerline{ \includegraphics[width=0.4 \textwidth]{icrc0916_fig02.eps}}
%\caption{\label{lba}Photograph of a LOFAR droop-dipole low-band antenna (LBA).}
%\end{figure}

Each station consists of either one or two high-band antenna (HBA) fields,
covering the frequency range $110$-$250$ MHz, and a low-band antenna (LBA) field,
covering the $10$-$80$ MHz range. The HBA fields consist of between $24$ and
$96$ tiles of $4 \times 4$ bowtie antennas, while the LBA fields
each consist of $96$ dual-polarization droop-dipole antennas, % (see Fig.~\ref{lba}),
48 of which can be used simultaneously (on the core and remote stations).
The LOFAR Cosmic Rays Key Science Project aims to use the LBAs to
detect atmospheric cosmic ray air showers in parallel with standard LOFAR
operations. 

\begin{table}
\begin{center}
\begin{tabular}{l | c c c c}
Station Type	& Number & LBA   & HBA 	\\
\hline
Core		& 24	&  $96^*$ 	& 2x24  \\
Remote		& 16	& $96^*$ 	& 48  \\
International	& 8	& $96\ $  	& 96  \\
\end{tabular}
\end{center}
\caption{Summary of the LOFAR station types. $^*$Only 48 can be used simultaneously.}
\label{summary_table}
\end{table}

As a fully digital telescope, LOFAR uses digital signal processing to focus
the signals from individual station elements, in contrast to traditional telescopes
which use geometrical optics (curved mirrors). Signals from individual elements
(LBA dipoles or HBA tiles) are digitized at $200$~MHz\footnote{A 160~MHz mode is also possible, but is used less frequently.}
with $12$-bit precision. The digitized signals are divided into $512$ frequency sub-bands,
up to $240$ of which can be added together with appropriate phase offsets to form up to eight
station beams viewing between $\sim2$ and $\sim1700$ square degrees each,
depending on the frequency selection and antenna configuration \cite{LOFAR}.
These beams are then sent to the LOFAR central processing hub in Groningen
for cross-correlation (imaging) or further beam-forming (e.g.~for pulsar searches).

In parallel to station beam-forming, the digitized time-domain signals from
each station element can also be analyzed with real-time logic and/or copied to
LOFAR's Transient Buffer Boards (TBBs), circular RAM buffers holding up to
$1.3$~seconds of data from each station element. Data from the TBBs can be accessed
and returned for processing without interfering with other observation modes.

The LOFAR Cosmic Rays Key Science Project is using this parallel capability
to search for atmospheric cosmic ray events. The LBAs individually
provide all-sky coverage, and their sensitivity peaks in the $30$-$80$ MHz range
where the maximum radio signal from extensive air showers (EAS) has been
observed \cite{falcke_nature}.  The LOFAR ``VHECR'' (very-high-energy cosmic ray) mode
utilizes the LBAs for a real-time search for EAS signatures, and its
implementation and operation is the subject of this contribution. The
``HECR'' (high-energy cosmic ray) mode, which trades sky coverage for increased sensitivity by using
station beams to search for EAS, will be developed in the near future, while the
``UHEP'' (ultra-high-energy particles) mode, by which beams from all
stations target the Moon to detect particle interactions in the Lunar
regolith, is the subject of another contribution \cite{uhep}. 

%------------------------------------------------------------

\subsection{Triggering and Data Acquisition}

In VHECR mode, LOFAR can trigger on cosmic rays in two ways: directly on
the radio air shower signal, or via an external trigger (e.g. from a
particle detector array).  Radio triggering is performed in two stages, and
the time-series signal around a pulse is stored for all channels once the
highest-level trigger is satisfied.

Our first-level trigger is based upon that developed for the LOPES experiment \cite{horneffer_trigger}.
Pulses are detected in real-time, by processing
the digitized radio signal of each polarization channel in an FPGA (field-programmable
gate array).  The FPGA performs a threshold search on the absolute values of
the voltage ADC samples using the criterion

\begin{equation}\label{threshold}
|x_i| > k\, \mu_i,
\end{equation}

\noindent where $\mu_i$ is a running average over the previous 4096 absolute values,
and $k$ is an integer threshold factor, typically set between
6 to 8.  If we assume the background signal to be Gaussian-distributed with
standard deviation $\sigma$, the mean absolute value is related to $\sigma$ as
\begin{equation}
\mu = \sqrt{\frac{2}{\pi}}\ \sigma \simeq 0.8\; \sigma, % sigma sqrt(2 / pi)
\end{equation}
\noindent and thus the threshold is typically a factor of 5 to 6 above the
time-varying noise level.  When this threshold is crossed, a message is
passed to the station's control PC, where the second-level trigger
operates.  The message contains a number of parameters characterizing the
pulse, including trigger time, pulse height, pulse width, and noise levels
before and after the pulse.  

The second-level trigger searches for coincidences between single-channel
triggers.  Typically, we require 16 to 32 out of 96 channels to trigger
within a 1 $\mu$s time window. For these events, we perform a real-time
direction estimate based on the first-level trigger times and the antenna
positions.  Although the uncertainty in time of
the trigger messages is about 2 to 3 samples (i.e. 10-15 ns), the estimated source direction is
accurate enough to distinguish between sources from the sky and interference
sources on the horizon.  The coincidence event rate is dominated by a few fixed radio-frequency
interference (RFI) sources ---
most commonly electric fences around nearby farm fields --- that emit pulsed radiation. Therefore, the second-level trigger
also uses this preliminary reconstruction to exclude triggers coming from
within {\bf $30^{\circ}$} of the horizon.

When a pulse passes all second-level trigger criteria, the
buffered time-series data for $0.5$ to $1$ ms around the pulse are stored for offline
analysis.  The relatively large size of this data window allows offline spectral cleaning
of narrow-band radio transmitters from the signal with a frequency resolution of 1 to 2 kHz. 

%\begin{figure}
%%\centerline{ \includegraphics[width=0.43 \textwidth]{LEDhires.eps}}
%\centerline{ \includegraphics[width=0.43 \textwidth]{icrc0916_fig03.eps}}
%\caption{\label{directions}Example of pulse arrival directions as calculated from trigger
%  arrival times. The center represents the zenith; the outer edge
%  corresponds to the horizon.  Shown are two fixed interference sources
%  and one pulse from a high elevation that would pass the trigger conditions. }
%\end{figure}

In addition to the radio-triggering mode, LOFAR can also be triggered by an
external particle detector array.  LORA (the LOFAR Radboud Air Shower Array) is an array
of 20 plastic scintillator detectors arranged in 5 stations within the
LOFAR superterp \cite{thoudam_lora}. Upon detection of an air shower by LORA, a trigger
can be sent to LOFAR to store the radio signal from all dipoles for further
analysis.  This can be used to provide an unambiguous cross-check that a
radio signal is from an EAS, and can also be used to reduce the detection
energy threshold, as described in Sec.~\ref{sec_threshold}.

\subsection{Data Analysis}

We process the second-level trigger data using an automated pipeline
structure, the first objective being to discriminate cosmic ray (CR) events
from other pulses. The following steps are performed:

\begin{compactenum}
\item data validation, e.g.~elimination of pulse trains or
  other noise;
\item matching of the event data with coincident triggers from particle detectors
  and/or other LOFAR stations;
\item spectral cleaning, i.e.~removal of narrow-band radio transmitters from
  the spectrum; and
\item reconstruction of the source position (direction and distance).
\end{compactenum}

To locate the source, we calculate the ``beam-formed'' time-series for a
set of directions and distances. Assuming a point source at a given
position, we correct for the relative time delays at each
antenna, using a spherical-wavefront approximation, and then sum the
delay-corrected signals from all antennas.
%(see Fig.~\ref{beamforming}). 
As the true source position can be identified
as the one yielding the maximum beam-formed pulse
power,
the best-fit direction can be efficiently found using a
downhill-simplex algorithm, using the second-level trigger's direction
estimate as a starting point for the search. The results of the pipeline can
then be used to determine key air shower event parameters such as energy
and position of shower maximum. 

\subsection{Energy Threshold and Estimated Event Rate}

\label{sec_threshold}

The detection energy threshold of the VHECR mode is determined at minimum by the
first-level trigger on an individual LBA dipole. An event with signal exactly equal to the threshold
will trigger on average half of the dipoles at station level (thus passing the
multi-channel trigger requirement), and will not be rejected as RFI unless it
is a highly-inclined event with zenith angle larger than $60^{\circ}$. Thus, the
second-level trigger restricts the sky to $\pi$ sr for real events.

The noise background against which cosmic ray signals must be detected is dominated by the
Galactic background emission, with effective temperature at wavelength
$\lambda$ given by
$T_{\rm sky} = 60 \pm 20 (\lambda / {\rm m})^{2.55}\ {\rm K}$ \cite{LOFAR}.
The sky temperature is in turn defined using the measured flux $I(\nu)$ via the Rayleigh-Jeans Law of $I(\nu) = 2 k_B T_{\rm sky} \lambda^{-2}$,
where $k_B$ is Boltzmann's constant.  The (sparse) outer LBAs have an
effective area $A_{\rm eff}$ that increases with wavelength: $A_{\rm eff} \approx \lambda^2/3$. As there
is strong RFI below $30$ MHz, we use a filter which limits our bandpass to approximately
$30$-$80$~MHz; after filtering, RFI typically contributes an additional 25\% to our system noise.
Assuming constant effective area over the observed solid angle $\Omega_{\rm sky}$ and
accounting for RFI, the noise
power $P_n$ received in a single polarization channel can be calculated by integrating over frequency $\nu$:
\begin{equation}
P_n = 1.25\ \Omega_{\rm sky} \int \frac{k_B \, T_{\rm sky}(\nu)}{\lambda^2} \, A_{\rm eff}(\nu) \, f(\nu) \, d\nu \label{noise}
\end{equation}
where $f(\nu)$ is the power response of the filter. The radio signal $V_{\rm eas}$ of an
extensive air shower is coherent and must be calculated in the voltage domain from its
intrinsic field strength $S(\nu)$ and angle $\theta_p$ of the receiver to the polarization vector:
\begin{equation}
V_{\rm eas} = \cos \theta_p \int S(\nu) \sqrt{f(\nu) A_{\rm eff}(\nu)} \, d\nu. \label{signal}
\end{equation}
This must sufficiently exceed the RMS noise voltage $V_{\rm rms}^2 = P_{n} Z_0$ in order to result in a trigger.
%\footnote{We use
%the free-space impedance $Z_0$ since antenna impedance affects both signal and
%noise equally.}.
Simulations of the expected signal $S(\nu)$ indicate an increased strength at
lower frequencies \cite{reas3_mgmr}; however, we use a empirical frequency-independent
parameterization based on LOPES measurements \cite{horneffer_param}.
%Note that since $f(\nu)$ decreases at low frequencies while both $T_{\rm sky}$ and $S(\nu)$
%increase at low frequencies, the integrands in both equations \ref{noise} and \ref{signal}
%will remain approximately constant.

Using a trigger threshold of $|V_{\rm eas}| > 5 |V_{\rm rms}|$, we find a minimum
energy threshold of $\sim5 \times 10^{16}$~eV.  However, 
the strength of the geomagnetic emission mechanism is strongly dependent on
the angle between the shower axis and the Earth's magnetic field, and a
typical shower geometry results in a threshold closer to $10^{17}$~eV.
Taking the station effective area to be equal to that of a circle with radius
given by the characteristic signal fall-off length of
$\sim 200$~m \cite{horneffer_param}, the total expected event rate for all $40$ core and
remote stations will be of the order of 1/hr.  The coincidence rate of
the superterp stations with the LORA particle detector is estimated to be
of order 1/day.

For LORA-triggered events (where the arrival direction is given by the particle detectors),
the radio threshold is given instead by the beam-formed data from all $48$
antennas. Thus the threshold for LORA events will be lower than radio-only mode by a factor of
between $\sqrt{48}$ and $\sqrt{96}$, depending on the relative strengths of the signal in
each of the polarization channels ($2$ per antenna). Therefore we expect a threshold of
$\sim 10^{16}$ eV over the central region covered by LORA,
and a corresponding event rate of order $2$/hr.

%------------------------------------------------------------

\subsection{First Cosmic Ray Events and Outlook} 

The first confirmed air shower radio pulses were observed with LOFAR in
June 2011, using LORA as a trigger.  The cosmic ray
nature of the pulses is confirmed by comparing independent directional
reconstructions by LOFAR and LORA, which typically agree to within a few
degrees. 

\begin{figure}
\centering
\includegraphics[width=0.45\textwidth]{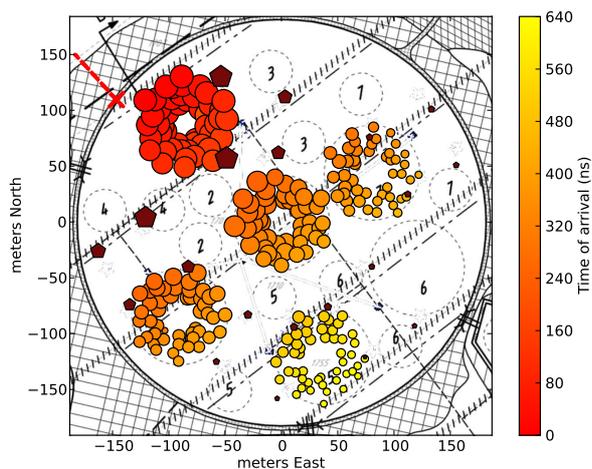}
\caption{\label{footprint} A cosmic ray air shower observed
  with five LOFAR stations (groups of circles) and LORA (pentagons).  The
  circle size represents the power of the radio signal, and the color scale
  indicates the arrival time.  The reconstructed shower core and arrival
  direction are shown by the cross and dotted line.} 
\end{figure}

\begin{figure}
\centering
\includegraphics[width=0.45\textwidth]{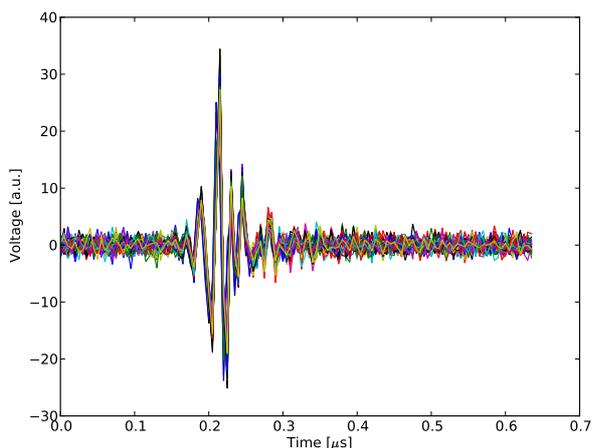}
\caption{\label{timeseries} Time-series air shower radio
  pulses from LOFAR antennas (superimposed), after correction for
  geometric arrival time delays.}
\end{figure}

\begin{figure}
\centering
\includegraphics[width=0.45\textwidth]{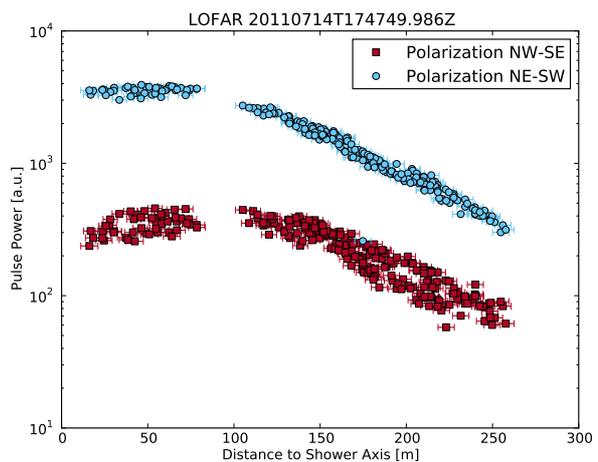}
\caption{\label{ldf} Radio pulse power as a function of distance to
  the shower axis, for a single air shower event seen in five LOFAR stations
  Both LBA polarizations are shown (aligned northeast-southwest and
  northwest-southeast).} 
\end{figure}

A sample air shower event is shown in
Figs.~\ref{footprint}, \ref{timeseries}, and \ref{ldf}.
 Fig.~\ref{footprint} shows the time delay and signal
 strength across the core stations.   Radio pulses from individual dipoles exhibit coherence after
correction for geometric delays (see Fig.~\ref{timeseries}).   A preliminary lateral distribution function of the same event, showing the
radio pulse power as a 
function of distance to the shower axis for the two LBA polarizations, is
shown in Fig.~\ref{ldf}.  The error bars are preliminary and are based on
average LORA reconstruction 
errors.  The relative normalization between the two polarizations does not
yet take into account differences in directivity, and a full calibration of the
radio signal into physical units (electric field) is still in progress.  

These data are, to our knowledge, the most densely instrumented measurements
of radio air shower emission to date.  Analysis of the lateral
distributions is underway and will allow detailed studies of the
polarization and falloff characteristics of the radio signal, including
effects from sub-dominant emission mechanisms.  Studies of
the wavefront shape are also planned.  

In addition to the LORA-triggered mode, optimization of the radio
self-trigger mode is continuing.  The improvements underway will allow a reduction of
deadtime as well as the veto of RFI sources from above the horizon,
such as those from airplanes.  The self-trigger mode will significantly
increase the detection area available for cosmic ray observation.

%------------------------------------------------------------

\clearpage
\end{document}